\documentclass[utf8]{frontiersinFPHY_FAMS} 

\setcitestyle{square} 
\usepackage{datetime, fmtcount, etoolbox, fcprefix}
\usepackage{url,hyperref,lineno,microtype,graphicx}
\usepackage[onehalfspacing]{setspace}
\usepackage{tabularx}
\usepackage{float}

\def\keyFont{\fontsize{8}{11}\helveticabold }
\def\firstAuthorLast{Vashishtha {et~al.}} 
\def\Authors{Nitin Vashishtha\,$^{1,*}$, Satabdwa Majumdar\,$^{1}$, Ritesh Patel\,$^{2}$, Vaibhav Pant\,$^{1}$, Dipankar Banerjee\,$^{1,3,4}$}
\def\Address{$^{1}$ Aryabhatta Research Institute of Observational Sciences, Nainital, UK 263002, India\\
$^{2}$ Southwest Research Institute, 1050 Walnut Street, Suite 300, Boulder, CO 80302, USA\\
$^{3}$ Indian Institute of Astrophysics, 2nd Block Koramangala, Bangalore 560034, India \\
$^{4}$ Center of Excellence in Space Science, IISER Kolkata, Kolkata 741246, India}
\def\corrAuthor{Nitin Vashishtha}

\def\corrEmail{nitin.vashishtha@aries.res.in}

\def\corrAuthor{Nitin Vashishtha}

\def\corrEmail{nitin.vashishtha@aries.res.in}
\usepackage{booktabs}
\begin{document}
\onecolumn
\firstpage{1}

\title[Imaging Cadence and CME Kinematics]{Exploring the Impact of Imaging Cadence on Inferring CME Kinematics} 

\author[\firstAuthorLast ]{\Authors} 
\address{} 
\correspondance{} 

\extraAuth{Satabdwa Majumdar \\ satabdwamajumdar@gmail.com}

\maketitle
\begin{abstract}
The kinematics of coronal mass ejections (CMEs) are crucial for understanding their initiation mechanism and predicting their impact on Earth and other planets. With most of the acceleration and deceleration occurring below 4 R$_\odot$, capturing this phase is vital to better understand their initiation mechanism. Furthermore, the kinematics of CMEs in the inner corona ($<$ 3 R$_\odot$) are closely related to their propagation in the outer corona and their eventual impact on Earth. Since the kinematics of CMEs are mainly probed using coronagraph data, it is crucial to investigate the impact of imaging cadence on the precision of data analysis and the conclusions drawn from it and also for determining the flexibility of designing observational campaigns with upcoming coronagraphs. This study investigates the impact of imaging cadence on the kinematics of ten CMEs observed by the K-Coronagraph of the Mauna Loa Solar Observatory. We manually track the CMEs using high cadence (15 s) white-light observations of K-Cor and vary the cadence as 30 s, 1 min, 2 min, and 5 min to study the impact of cadence on the kinematics. We also employed the bootstrapping method to estimate the confidence interval of the fitting parameters. Our results indicate that the average velocity of the CMEs does not have a high dependence on the imaging cadence, while the average acceleration shows significant dependence on the same, with the confidence interval showing significant shifts for the average acceleration for different cadences. The impact of degraded cadence is also seen in the estimation of the time of onset of acceleration. We further find that it is difficult to find an optimum cadence to study all CMEs, as it is also influenced by the pixel resolution of the instrument and the speed of the CME. However, except for very slow CMEs (speeds less than 300 Kms$^{-1}$), our results indicate a cadence of 1 min to be reasonable for the study of their kinematics. The results of this work will be important in the planning of observational campaigns for the existing and upcoming missions that will observe the inner corona.
\section{}
\tiny
 \keyFont{ \section{Keywords:} Sun, Corona, Coronal Mass Ejections (CMEs), CME Kinematics, Bootstrap} 
\end{abstract}

\section{Introduction}
\label{sec-intro}
Coronal mass ejections (CMEs) are explosive large scale eruptions of magnetic field and plasma from the Sun's corona into the heliosphere \citep{hundhausen1984,Gopalswamy2004,yashiro2004,Webb2012}.
They can travel at speeds ranging from a few hundred to a few thousand km s$^{-1}$ and can accelerate at a rate ranging from a few tens to a few 10$^{4}$ m s$^{-1}$ \citep{Webb2012}. CMEs are considered one of the primary drivers of space weather hazards on Earth \citep{hapgood} since they have the potential to generate shock waves and geomagnetic storms, which can result in technological damage on Earth \citep{gosling1993,Schwenn2006,Pulkkinen2007}. Therefore, it is crucial to comprehend the kinematics of CMEs from the inner corona all the way to Earth.\par 
The CME propagation is impacted by a dynamic interaction between the following forces: the gravitational force, the Lorentz force, the pressure gradient force, and the force of viscous drag due to background solar wind \citep{Wood_1999,Zhang_2001,VRSNAK2006431,Vršnak2007,sachdeva,Majumdar_2020,Lin}. These forces play a huge role in dictating the kinematic profile of the CMEs. The three-part kinematic profile \citep{Zhang_2006} is commonly observed in most CMEs as a result of the interplay between these forces. The first phase of CMEs is a slow rise \citep{Cheng_2020}, followed by an impulsive phase characterized by a rapid increase in speed \citep{Bein_2011,Gallagher_2003,Temmer_2008,Joshi_2011,Cheng_2020,Majumdar_2020,Patel2021}. \cite{Gui2011} suggested that the impulsiveness of CMEs is observed below a height of 1.5 R$_{\odot}$. Finally, a phase with little to no acceleration where they experience the drag force due to background solar wind \citep{gopalswamy2000,Moon_2002,Vršnak2002,Cargill2004,Borgazzi,Majumdar_2021}
. Although some progress has been made in understanding the third phase of the CME kinematic profile \cite{Sachdeva2017}, the slow rise phase and the impulsive acceleration phase \citep{Cheng_2020}, which are the first two phases of the profile, remain inadequately understood, owing partly to a lack of uninterrupted observation in the inner corona \cite{Bein_2011}. The first two phases, however, hold immense significance as they provide valuable insights into the initiation mechanisms of CMEs.  Capturing the curvature in the kinematic profile is critical since it serves as a good indicator of the underlying mechanism that triggers CMEs (see \cite{Mierla2013}) and on estimating the height of shock formations \citep{majumdar_2021b}. Given the significance of the curvature in the kinematic profile, it is imperative to ensure that the observing instrument has a suitable cadence to capture this curvature accurately.
Recently \cite{Majumdar_2020} established a strong correlation between the magnitude and duration of the actual 3D acceleration. However, a critical consideration in their analysis was the necessity for the instrument to have the capability of capturing the entire acceleration duration, which requires an instrument with a good field of view (FOV) and high cadence. K-Coronagraph \citep{10.1117/12.926511} is one such ground based instrument which observes the CMEs in the low coronal heights (FOV of 1.05-3 R$_{\odot}$) with a minimum cadence of 15 s, which has recently been exploited for the first ever combined space and ground based stereoscopy in the inner corona by \cite{Majumdar_2022}.\par
The accurate prediction of CME arrival time relies not only on understanding the dynamics of CMEs in the heliosphere \citep{Amerstorfer2021-ur,TEMMER2023} but also on considering the impulsive acceleration and the constant speed third phase, which both play significant roles. The impulsive acceleration phase that occurs within the inner corona is closely linked to the subsequent constant-speed third phase. \cite{Majumdar_2021} recently highlighted the close relationship between kinematic parameters in the inner corona (3 R$_{\odot}$) and those in the outer corona and how the latter can be estimated from the former. However, the accuracy of estimating these parameters can be enhanced through improvements in cadence. Apart from that, accurately knowing the arrival time of CMEs at 1 AU is crucial for predicting space weather, as severe geomagnetic storms are caused by CMEs. Various models have been used to predict CME arrival times, including empirical models \citep{Gopalswamy2001,Paouris2017}, shock propagation models \citep{Dryer2001,Zhao_2016,Takahashi_2017}, Drag-Based models (DBM) \citep{Vršnak2013,Hess_2014,Hess_2015,zic_2015,Dumbovic}, and numerical magnetohydrodynamics (MHD) models \citep{Mikic1999,Odstrcil2004,wu2011,pomoell2018}. Some of these models use kinematic parameters of CMEs near the sun as their input parameters for CME arrival time prediction. \cite{kay2018} demonstrated the impact of uncertainties in the initial input parameters of a CME on the accuracy of its arrival time prediction. \cite{byrne} studied the effect of the sampling cadence on deriving the kinematics from a simulated constant acceleration profile of a coronal wave. They found that higher-cadence data provides a clear and accurate representation of the true kinematic profile, allowing for precise estimations of the velocity and acceleration. Therefore, it will be interesting to investigate the effect of sampling cadence on CME kinematics for different CMEs observed from different instruments. Averaging to longer cadence windows can be seen as applying a smoothing filter on the CME kinematic profile. As a result, there are compelling a priori reasons to expect a change in the speed and accelerations inferred from CME kinematic profile. Hence, it is essential to constrain the fitting parameters of the kinematic profile, which is, in turn, influenced by the cadence of the instruments used in estimating those parameters. 
This study thus aims to investigate the impact of imaging cadence on inferring the kinematics of the CMEs in the inner corona using K-Coronagraph. As a ground-based instrument, the K-Cor coronagraph is susceptible to fluctuations in weather conditions that may have an impact on the observations \citep{Thompson}. With a spatial resolution of 11.3 arcseconds and a minimum cadence of 15 s, K-Cor provides valuable observations. However, it will be intriguing to explore the implications of inferring kinematics from the capabilities of the existing and upcoming space-based instruments. For instance, Metis on board Solar Orbiter \citep[Metis; ][]{metis}, Visible Emission Line Coronagraph \citep[VELC; ][]{singh2011,Singh2013,raghavendra} on board Aditya-L1 \citep{seetha}, and The Sun Coronal Ejection Tracker \citep[SunCET;][]{Suncet} offer enhanced spatial and temporal resolutions. VELC provides 5 arcseconds spatial resolution, while Metis achieves 5.6 arcseconds (at its closest perihelion of 0.28 AU) with a minimum cadence of 1 sec. Hence, our findings are anticipated to offer valuable insights into the challenges and prospects that await us in understanding the kinematic profile of CMEs by working on the data from the above-mentioned instruments.
This paper is organised into different sections. Section 2 presents the data source used and the methodology adopted, while Section 3 provides our findings. Finally, in Section 4, we summarize the key conclusions from our study. 
\section{Data and Method}
\label{sec-obs}
\subsection{Data Source}
In this study, we utilized data obtained from the K-Cor coronagraph, which is a ground-based instrument located at Mauna Loa Solar Observatory (MLSO). The K-Cor instrument provides a field of view ranging from 1.05 to 3 solar radii. We used level 2.0 Normalizing-Radial-Graded Filter \citep[NRGF;][]{Morgan2006} processed data with cadences of 15 s and 2 min. We then created 30 s, 1 min, and 5 min average image data sets using the 15 s cadence data. We created these datasets by taking the arithmetic mean of the 15 s cadence images. It should be noted that the images so produced after averaging the 15 s cadence images will have some effect of the motion blur because of the propagation of the CMEs in the time interval between the first and last images considered for the averaging. The effect of the motion blur will be different on different data-set images because of different time intervals, and that would make it tricky to track features inside the CME leading front. However, it is worth noting that single viewpoint tracking of the CME front can invite many uncertainties, as pointed out by \citep{barnard2015}, and in this regard, advanced methods of characterizing the CME fronts \citep{barnard2017} can be incorporated in future studies. To enhance the relevant features and eliminate the static features and contributions from the F-corona, we subtracted an image captured prior to the onset of the CME, creating a base difference image for each cadence, including 30 s, 1 min, 2 min, and 5 min. It should be noted that the base image prior to the onset of the CME for a particular cadence is chosen from the same cadence dataset.
\subsection{Event Selection}
The K-Cor coronagraph, being a ground-based instrument, is subject to the effects of varying weather conditions that can potentially impact the quality of its data. Localized effects, such as wind-blown dust and insects entering the telescope's field of view, can degrade the observing conditions. Wind can also disturb the stability of the telescope's pointing. Furthermore, the bright sky background presented an additional challenge during our analysis \citep{Thompson}. 10 CME events occurring between November 2014 and September 2022 were selected for our study. Our selection criteria were based on identifying CMEs with a bright and distinct front within the field of view (FOV) of the K-Cor instrument while excluding those with large deflections that may introduce errors in our tracking analysis. We also identified the source region of these events. CMEs that occurred on May 07, 2021, and January 10, 2022, are linked to active regions (ARs) and the remaining 8 CMEs are connected to active prominence eruptions (APs). In a study conducted by \citep{Majumdar_2021}, it was demonstrated that there are distinct differences in the kinematic properties of coronal mass ejections (CMEs) originating from active regions (ARs) and active prominence eruptions (APs). Therefore, it becomes crucial to examine the impact of cadence on both types of CMEs. Please note that the primary objective of this study is to present a proof of concept regarding the impact of imaging cadence on inferred kinematics. To enhance the statistical significance of the results, it is recommended to expand the study to include a larger and more diverse subset of CMEs. This could be achieved by incorporating CMEs originating from various source regions, as facilitated by the recently published CME source region catalogue by \cite{majumdar2023cme}.
\subsection{Method}
\begin{figure}[!ht]
    \centering
    \begin{minipage}[!ht]{0.45\textwidth}
        \centering
        \includegraphics[width=\textwidth]{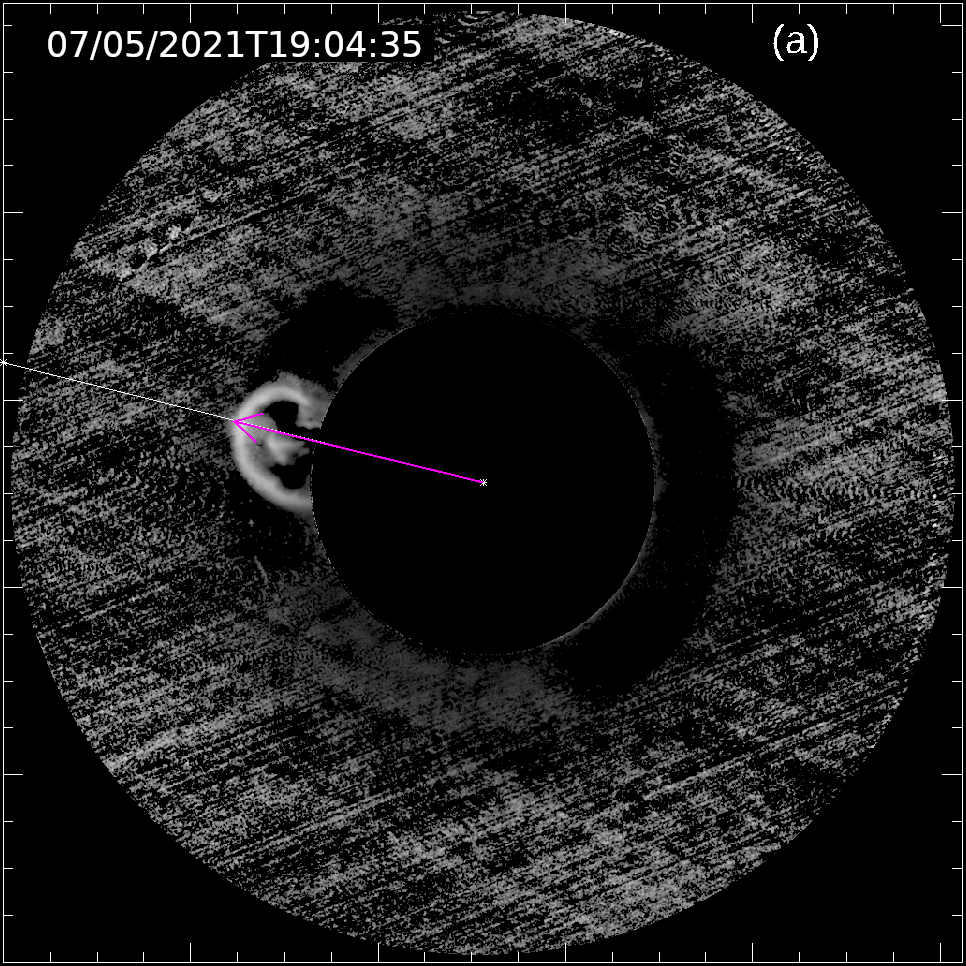}
    \end{minipage}
    \hfill
    \begin{minipage}[!ht]{0.45\textwidth}
        \centering
        \includegraphics[width=\textwidth]{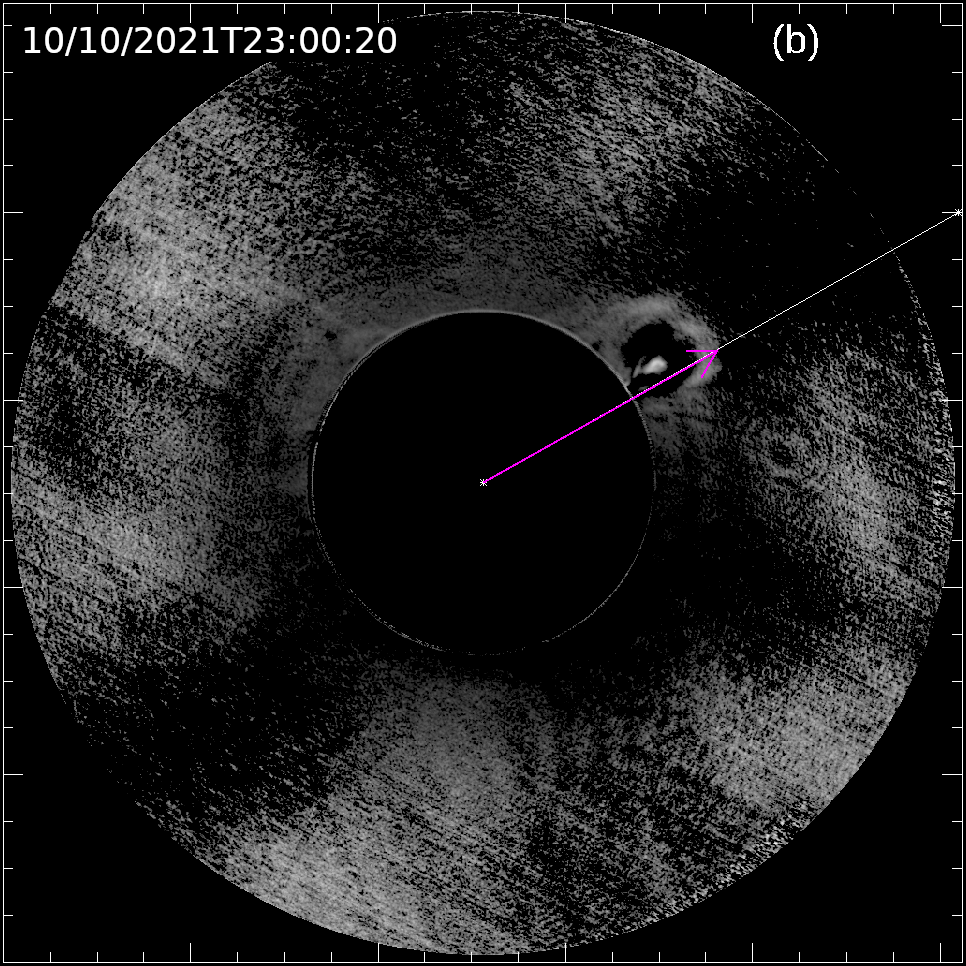}
    \end{minipage}
    \caption{Difference Images of CMEs that occurred on (a) 07/05/2021, and (b) 10/10/2021 in the K-Cor field of view. The white line represents the angle for CME tracking, while the magenta arrow indicates the position of the leading edge, and its length represents the height of the leading edge.}
    \label{fig:tracking}
\end{figure}

To track the CME front, we carefully selected the part of the leading edge that remained visible throughout the entire field of view of K-Cor in all images. Next, we drew a reference line passing through the selected part of the CME from the centre of the sun. Along this particular angle, we marked the positions of the leading edge to track the CME's motion(Figure \ref{fig:tracking}). This process also allows us to minimize the possibility of tracking different parts of the CME leading edge at different times. Since, in this work, we are interested in the radial kinematics of CMEs and the influence of image cadence on the same, we ignore deflections experienced (if any) by the CMEs.\par

\begin{figure}[!ht]
    \centering        \includegraphics[width=\textwidth]{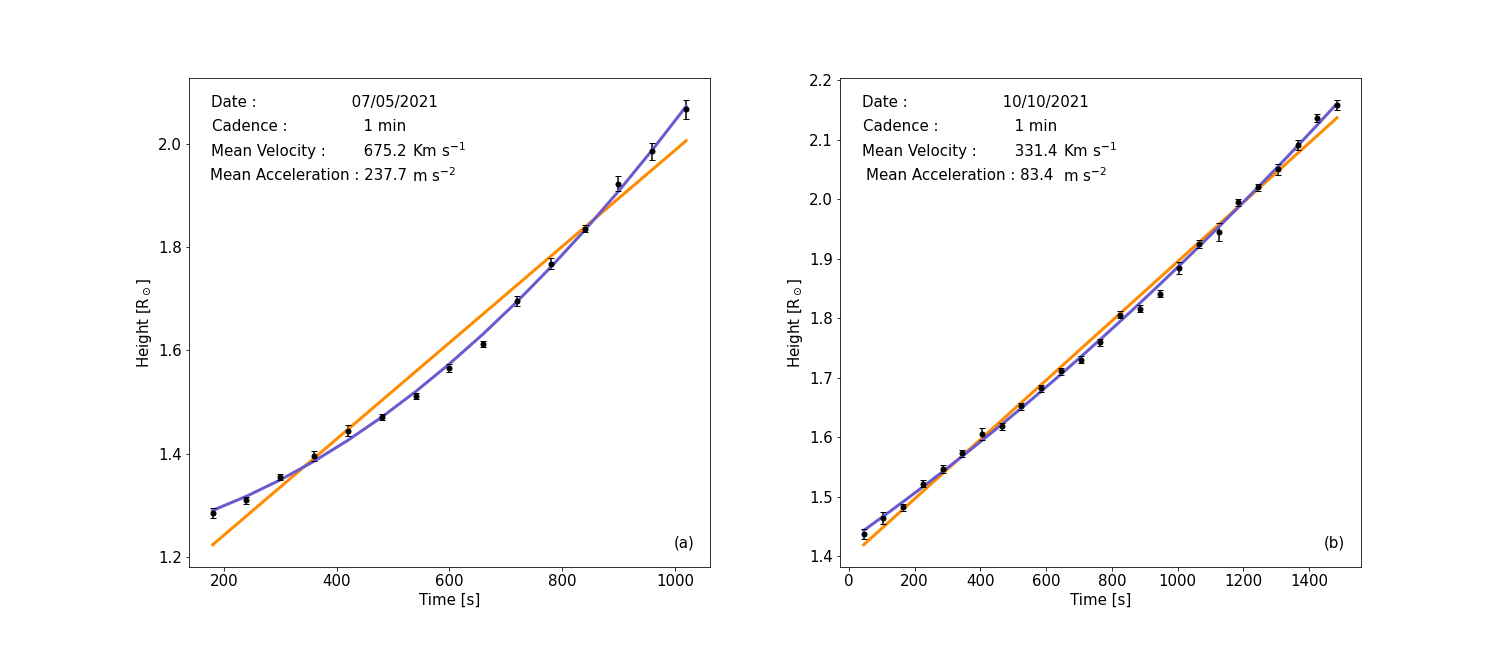}
    \caption{Height time plot of CME occurred on (a) 07/05/2021, (b) 10/10/2021 for 1 min cadence. The linear fit is represented by the orange line, while the quadratic fit is represented by the blue curve. The estimated mean velocity and acceleration are also indicated.}
    \label{fig:fiting}
\end{figure}
We use first and second order polynomials to fit the height-time profiles (Figure \ref{fig:fiting}) and determine the average velocity and acceleration for the selected CMEs, respectively. When attempting to determine the average velocity and acceleration and subsequently assessing the accuracy of these parameters, a small sample size presents an immediate limitation. Therefore, to approximate the behaviour of the true distribution, we have used bootstrapping technique \citep{Efron1979,Efron1994,chernick1999boostrap}, which is based on the resampling method. Bootstrapping is a resampling technique used to create an approximation of the underlying distribution by repeatedly sampling the data. This process involves generating multiple samples, with replacements, from the original data. Each sample serves as a maximum likelihood estimator, allowing us to understand the statistical characteristics of the data better. Following are the steps for the implementation of the bootstrapping technique:\\
\textit{Step 1.} First, we obtain an initial fit to the data, which gives us the model fit with its corresponding fitting parameters.\\
\textit{Step 2.} Next, we calculate the residuals of the fit as the difference between the original data and the model fit.\\
\textit{Step 3.} To perform bootstrapping, we randomly resample the residuals with replacement to create a new set of residuals without removing any data points from the main data set.\\
\textit{Step 4.} With the new set of residuals, we generated a new data set and fit it with the model in order to get the new fitting parameters.\\
\textit{Step 5.} Steps 3 and 4 are repeated many times to generate a large number of bootstrapped datasets.\\
\textit{Step 6.} Finally, we calculate confidence intervals on the model parameters by analyzing the resulting distributions obtained from the repeated fits. These confidence intervals provide a measure of uncertainty in the estimated parameters.\\
In this work, during the bootstrapping process, we generated 10,000 datasets for each cadence for each event. We have bootstrapped both the linear and the quadratic fitting, which allows us to calculate the 95\% confidence intervals of the average velocity and average acceleration. However, a limitation of this approach is that for the case of less number of data points (as in the case for 5 min cadence), the bootstrapping technique will end up generating many degenerate samples, which will affect the inferred confidence intervals. \par
In order to get the kinematic profiles of CMEs, we have used the Savitzky-Golay filter \citep{Savitzky} to fit the height-time plots and generate corresponding velocity and acceleration profiles. In their study, \citep{byrne} demonstrated the benefits of utilizing the Savitzky-Golay filter over the conventional 3-point Lagrangian method. The Savitzky-Golay filter can provide a better smoothing of small-scale fluctuations while preserving the underlying kinematic profile, especially for the high cadence height time plots. These profiles are obtained by taking the first- and second-order numerical derivatives of the height-time data. We fit the data set with an average window size of 5 to 15. We selected the window size based on the number of data points in each dataset. If a dataset had a smaller number of data points, then we reduced the window size accordingly. A window size of 5 implies that we are considering 2 neighbouring data points on each side of the data point during the smoothing process. To take care of the endpoints, we handle the boundary conditions using mirror padding, ensuring that the endpoints receive the same level of smoothing as the rest of the data. This avoids introducing artificial effects at the edges of the data. The filter was applied with a selected polynomial order of 3 for the height time plot and 2 for the velocity time plot. The polynomial order of 3 for the height-time data has been chosen, keeping in mind that the CMEs experience a non-zero acceleration in these lower heights, and this 3rd order height-time polynomial incorporation ensures a linear acceleration profile for the CMEs. We realise that even a linear acceleration profile is not completely true, especially at these lower heights, but it would provide a rough estimate of the average trend of acceleration that the CME experienced. This suggests that we have made the assumption that the acceleration remains constant and follows a linear trend within the selected window size. Higher polynomial orders result in more aggressive smoothing of the data. A higher-order polynomial can better fit and remove small-scale noise and fluctuations in the data, leading to a smoother output. However, it is essential to note that increasing the polynomial order excessively can lead to overfitting the data, which means the filter may fit the noise or random fluctuations in the data instead of capturing the true underlying pattern. This can lead to a loss of important features and introduce artefacts in the filtered signal. The bootstrapping process on the Savitzky-Golay filter allowed us to calculate the median and the interquartile range (IQR). The interquartile range (IQR) is a measure of variability in a dataset, calculated as the difference between the upper and lower quartiles, where the latter two correspond to the 25th and 75th percentiles.

\section{Results}
\label{sec-result}
\begin{figure}[!ht]
    \centering
    \begin{minipage}[!ht]{0.48\textwidth}
        \centering
        \includegraphics[width=0.95\textwidth,height=20cm]{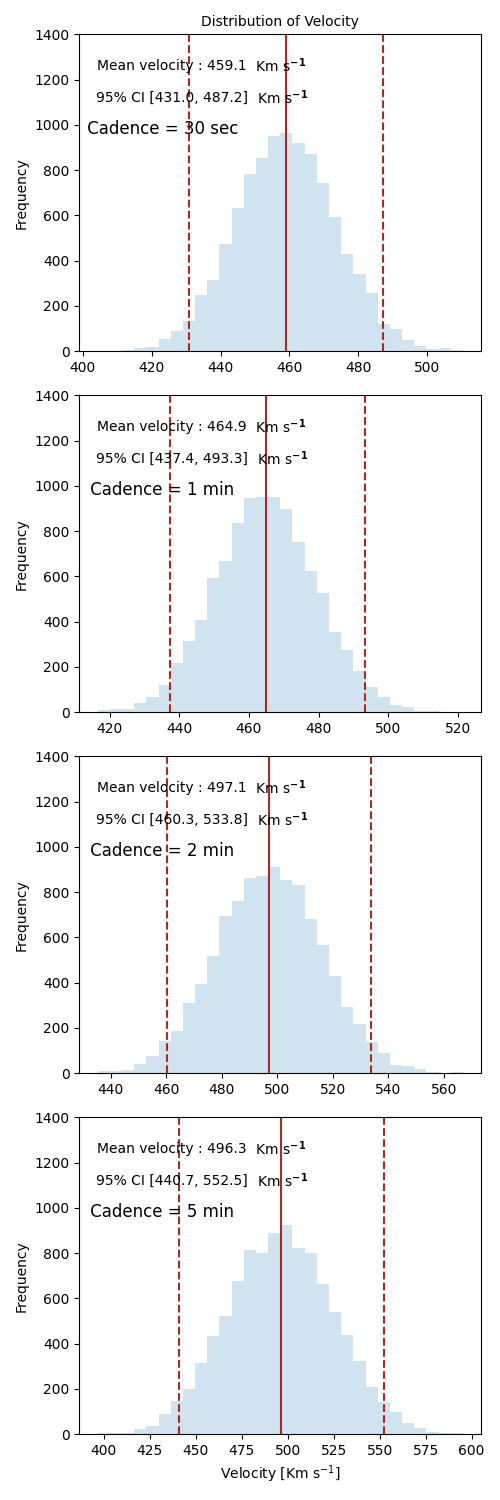}
    \end{minipage}
    \hfill
    \begin{minipage}[!ht]{0.48\textwidth}
        \centering
        \includegraphics[width=0.95\textwidth,height=20cm]{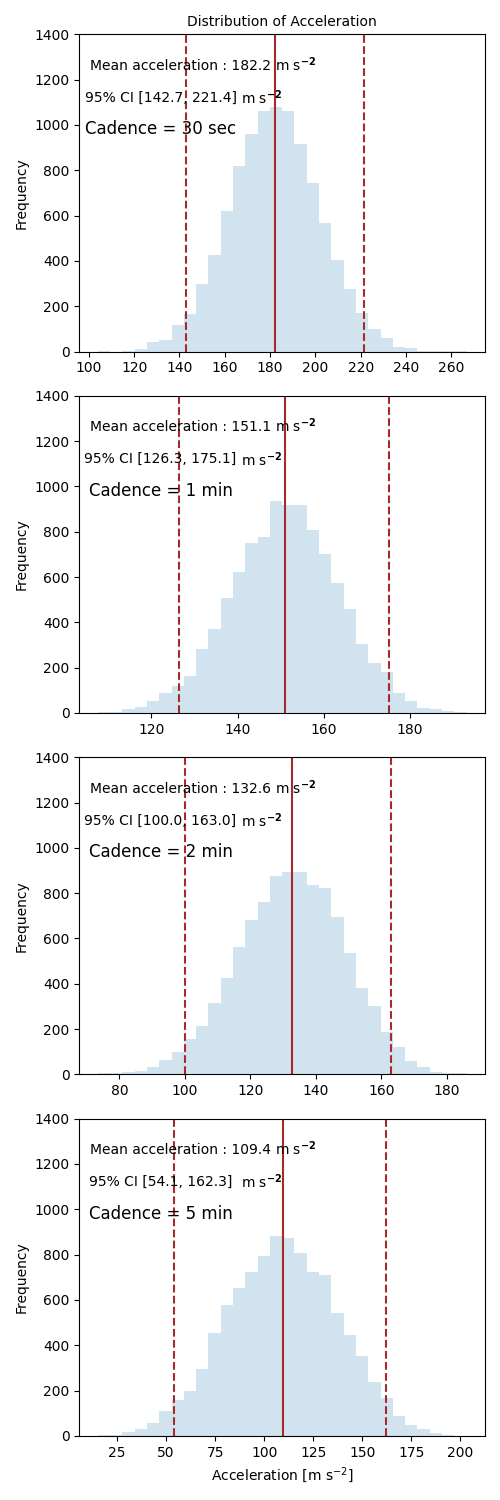}
    \end{minipage}
    \caption{Histograms depicting the distribution of velocity (left column) and acceleration (right column) values obtained using the linear fit and quadratic fit method, respectively, with the application of the bootstrapping technique for the CME occurred on June 10, 2021. The dashed and solid line represents the 95\% confidence interval and mean of the distribution, respectively.}
    \label{fig:boot_vel_acc}
\end{figure}
\begin{figure}[!ht]
    \centering
    \setcounter{subfigure}{0} 
    \begin{minipage}[ht]{0.45\textwidth}
        \centering
        \includegraphics[width=\textwidth]{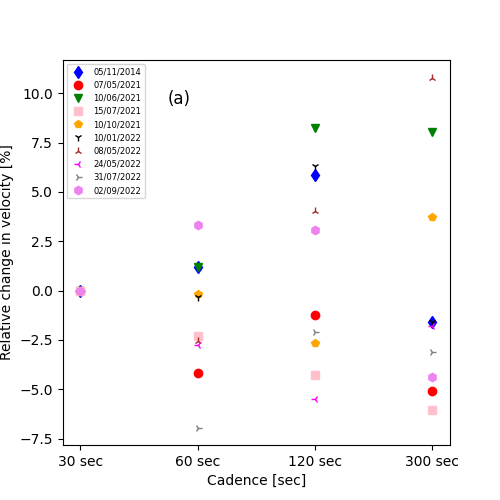}
    \end{minipage}
    \hfill
    \setcounter{subfigure}{1} 
    \begin{minipage}[ht]{0.45\textwidth}
        \centering
        \includegraphics[width=\textwidth]{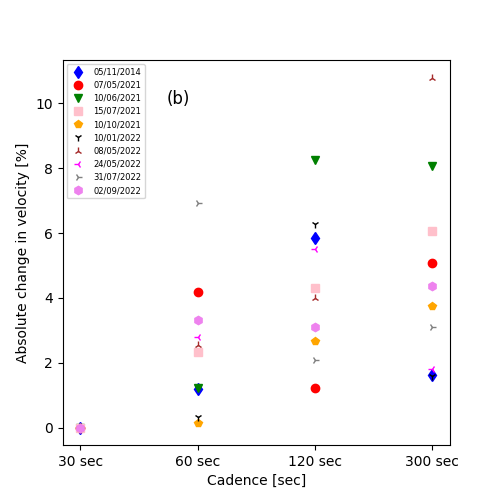}
    \end{minipage}
\caption{The (a) relative and (b) absolute changes in the velocity of each CME with respect to their 30 s cadence values.}
\label{fig:abs_rel_vel}
\end{figure}

Table \ref{table:1} shows the mean velocity values for each cadence of all the events obtained from the bootstrapped distribution. The table also includes the 95\% confidence interval of the bootstrapped distribution of the velocity. The date of the event is presented in the first column of the table. Figure \ref{fig:boot_vel_acc} represents the bootstrapped distribution of the velocity for all the cadences for the CME on June 10, 2021. Figure \ref{fig:boot_vel_acc} is just a representative example, as the plots for the same on all the other studied CMEs yield similar results, and besides, the idea we wish to communicate through this figure could be communicated with a single example as we have shown here. It should be noted that no preference has been considered in choosing this particular event (see supplementary material for more examples). We found that the mean of the distribution and the 95\% confidence interval range of the CME velocity distribution shift towards slightly higher velocities with an increase in the cadence, thus showing the sensitivity of sampling cadence. To add quantitative rigour to our analysis and assess the significance of changes in the 95\% confidence interval, we employ the two-sample Kolmogorov-Smirnov (KS) test on the velocity and acceleration distributions for various cadences. The KS test allows us to calculate the KS statistic along with the p-value, which measures the maximum vertical difference between the cumulative distribution functions (CDFs) of two compared samples and the probability of obtaining the observed KS statistic, respectively. During the KS test, we evaluated the null hypothesis, assuming that the two distributions are identical, while the alternative hypothesis suggests they are different. A larger KS statistic value and smaller p-value suggest evidence against the null hypothesis in favour of the alternative, indicating a greater dissimilarity between the two CDFs. In simpler terms, a higher KS statistic suggests stronger differences between the two distributions being compared. In Table \ref{table:3}, we present the results of the two-sample KS test between the velocity distributions for different cadences. The off-diagonal elements of the table show the KS statistic values between corresponding distributions. We observe that the KS statistic values are notably larger between 30 seconds and higher cadence. The p-value was found to be less than 0.05 for all the comparisons made, which indicates significant differences in the distributions of the two samples. The average speed of a CME is a very commonly used parameter, not only for kinematic diagnostics but also for space weather forecasting, and hence it is important to study the impact of change in cadence on the estimate of the average speeds. Thus, we calculated the relative change in percentile in the velocity values with respect to their 30 s cadence value for all the events and for all the cadences
(Figure \ref{fig:abs_rel_vel}(a)). We also plot the absolute of this relative change in velocity with respect to their 30 s cadence value for all the events for all the cadences (Figure \ref{fig:abs_rel_vel}(b)). Each data point is colour-coded to reflect a specific event. We observed that for certain CMEs, the velocity increases with an increase in the cadence, while for others, it decreases with the increase in cadence (Figure \ref{fig:abs_rel_vel}(a)). We also noticed that the change in velocity relative to the 30 s cadence was within 12\% of the value of the 30 s cadence (Figure \ref{fig:abs_rel_vel}(b)). This could be attributed to the assumption that these values are based on linear fitting, which assumes that there is no curvature in the CME kinematic profile. The visibility of the curvature in the kinematic profile depends heavily on the sampling of cadence, and thus, it is expected that the average velocity from linear fit would not show a genuine dependence on the cadence.\par
\begin{figure}[!ht]
    \centering
    \setcounter{subfigure}{0} 
    \begin{minipage}[ht]{0.45\textwidth}
        \centering
        \includegraphics[width=\textwidth]{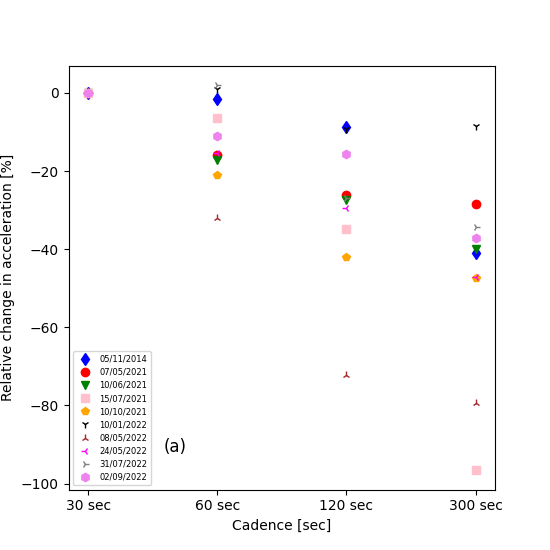}
    \end{minipage}
    \hfill
    \setcounter{subfigure}{1} 
    \begin{minipage}[ht]{0.45\textwidth}
        \centering
        \includegraphics[width=\textwidth]{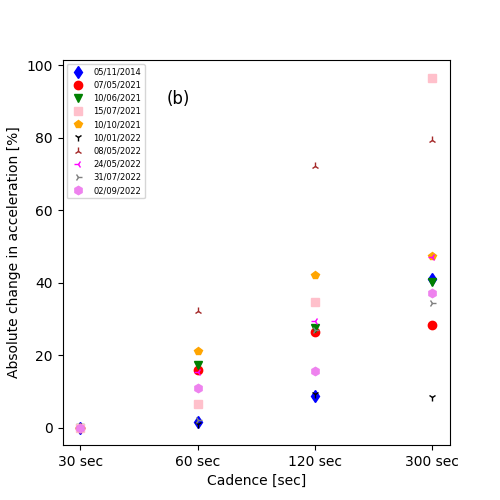}
    \end{minipage}
\caption{The (a) relative and (b) absolute changes in the acceleration of each CME with respect to their 30 s cadence values.}
\label{fig:abs_rel_acc}
\end{figure}

Similarly, Table \ref{table:2} shows the acceleration values obtained from the bootstrapped quadratic fit for each cadence in all events. Figure \ref{fig:boot_vel_acc} illustrates the distribution of the acceleration obtained through bootstrapping for all the cadences of the CME that occurred on June 10, 2021. We find that the confidence interval undergoes significant change, with a general trend of being shifted towards lesser acceleration values with an increase in cadence. In Table \ref{table:4}, we present the results of the two-sample KS test between these samples of acceleration for different cadences. The off-diagonal elements of the table show the KS statistic values between corresponding distributions. We noticed that the KS statistic values are even larger than compared to the KS statistic values for velocity distribution between 30 seconds and higher cadence, signifying significant differences in the distributions of the two samples. This shows the importance of having a good cadence to capture the curvature in the height-time plot and hence the acceleration phases experienced by the CME. We performed a similar analysis to obtain the relative and absolute values of average acceleration for each cadence relative to their 30 s cadence value. We then plotted these values against the cadence (see Figure \ref{fig:abs_rel_acc}). We again find significant changes in acceleration for different cadences. We noticed that the acceleration tends to decrease as we increase the cadence (Figure \ref{fig:abs_rel_acc}(a)). This clears hints at the fact that acceleration is a measure of the curvature in the kinematic profile, and as the cadence is degraded, less and less curvature is captured by the height-time data. We see a huge change in the value of acceleration with the cadence, particularly for CME occurred on 15 July 2021. We observe that some of the CMEs show a change in average acceleration of more than 70\% with respect to their 30 s cadence value (Figure \ref{fig:abs_rel_acc}(b)). Thus, it clearly shows the importance of image cadence in inferring the accelerations (and hence the magnitude of the net driving force) experienced by a CME. \par

\begin{table}[!ht]
\centering
\caption{Bootstrapped velocity and its 95\% confidence interval obtained from linear fitting to height-time plot for different CMEs corresponding to different cadences.}
{\footnotesize
\begin{tabular*}{\textwidth}{
  @{\extracolsep{\fill}}
  c
  c
  c
  c
  c
  @{}
}
\toprule
 \multicolumn{1}{c}{\textbf{Date}} & \multicolumn{4}{c}{\textbf{Velocity [95 \% Confidence Interval] (Km s$^{-1}$)}}\\\cline{1-5}\addlinespace
  & \multicolumn{4}{c}{\textbf{Cadence}}\\ \cline{2-5}\addlinespace
  & \textbf{30 s} & \textbf{1 min} & \textbf{2 min} & \textbf{5 min} \\
\midrule
   November 05, 2014       & 385.3 [370.3, 400.6]        & 389.9 [368.2, 412.0]     & 407.9 [379.2, 436.5]       &  379.0 [351.6, 406.4]  \\
\addlinespace
May 07, 2021          &   704.8 [673.8, 735.9]         &  675.1 [635.7, 714.6]        &  696.4 [652.1, 741.4]         & 668.0 [590.2, 746.8] \\
\addlinespace
June 10, 2021             &  459.0 [431.0, 487.0]          &  464.8 [436.3, 492.4]        &    496.8 [460.3, 534.0]        & 496.2 [439.5, 552.2]       \\
\addlinespace 
July 15, 2021   &   1035.4 [1021.9, 1049.1]        &     1011.8 [992.7, 1032.1]    & 991.8 [951.7, 1031.5] &      972.6 [960.0, 984.6]        \\
\addlinespace 
October 10, 2021 &    346.3 [334.1, 358.9]       &    346.0 [332.1, 360.3]     & 337.3 [322.3, 353.2] &     358.8 [337.1, 381.2]         \\
\addlinespace 
January 10, 2022 &  451.8 [435.3, 468.4]         &  450.9 [424.5, 478.0]       & 479.4 [446.9, 512.7] &   445.2 [392.1, 496.2]           \\
\addlinespace 
May 08, 2022 &   276.4 [264.3, 288.0]        &  269.6 [259.2, 279.9]       & 287.6 [280.1, 295.2]  & 306.6 [294.7, 319.0]             \\
\addlinespace 
May 24, 2022 & 405.9 [393.0, 418.4]          &    394.9 [380.8, 409.5]     & 384.0 [366.1, 400.0] &  398.5 [371.8, 425.1]            \\
\addlinespace 
July 31, 2022 &  648.4 [613.3, 682.3]         &   602.7 [551.0, 663.6]      & 635.6 [576.2, 689.4] &    628.8 [553.5, 703.4]          \\
\addlinespace 
September 02, 2022 &  354.9 [340.0, 370.2]         &   366.9 [345.3, 388.6]      & 365.4 [336.1, 394.2]  &   339.5 [309.2, 369.6]           \\
\bottomrule
\end{tabular*}
}
\label{table:1}
\end{table}
\begin{table}[!ht]
\centering
\caption{Two sample KS test statistic values for the bootstrapped velocity distribution for all the cadences of CME occurred on June 10, 2021.}
{\footnotesize
\begin{tabular*}{\textwidth}{
  @{\extracolsep{\fill}}
  l
  c
  c
  c
  c
  @{}
}
\toprule
  & \multicolumn{4}{c}{KS statistic}\\ \cline{1-5}\addlinespace
 \multicolumn{1}{c}{Cadence} & 30 s & 1 min & 2 min & 5 min \\
\cline{2-5}
   \multicolumn{1}{c}{30 s}       &     \multicolumn{1}{|c}{-}    &  0.16    &  0.75       &   0.64 \\
\multicolumn{1}{c}{1 min}          &     \multicolumn{1}{|c}{0.16}       &  - & 0.67           & 0.57  \\
\multicolumn{1}{c}{2 min}             &  \multicolumn{1}{|c}{0.75 }          & 0.67          &  -          &  0.11       \\ 
\multicolumn{1}{c}{5 min}   &  \multicolumn{1}{|c}{0.64 }         & 0.57        & 0.11  &  -            \\
\bottomrule
\end{tabular*}
}
\label{table:3}
\end{table}
\begin{table}[!ht]
\centering
\caption{Bootstrapped acceleration and its 95\% confidence interval obtained from quadratic fitting to height-time plot for different CMEs corresponding to different cadences.}
{\footnotesize
\begin{tabular*}{\textwidth}{
  @{\extracolsep{\fill}}
  l
  c
  c
  c
  c
  @{}
}
\toprule
 \multicolumn{1}{c}{\textbf{Date}} & \multicolumn{4}{c}{\textbf{Acceleration [95 \% Confidence Interval] (m s$^{-2}$)}}\\\cline{1-5}\addlinespace
  & \multicolumn{4}{c}{Cadence}\\ \cline{1-5}\addlinespace
  & 30 s & 1 min & 2 min & 5 min \\
\midrule
   November 05, 2014       &     129.2 [118.1, 139.4]     &  127.3 [106.3, 147.1]    &  118.3 [86.8, 149.5]      &   76.4 [58.8, 95.3] \\
\addlinespace
May 07, 2021          &     282.7 [243.8, 318.1]       &  236.9 [192.8, 275.8]        & 208.8 [175.9, 240.1]          & 203.8 [125.0, 281.3] \\
\addlinespace
June 10, 2021             &  182.6 [143.7, 220.5]          & 150.9 [126.4, 174.4]         &  132.2 [99.3, 163.0]          &  109.0 [54.9, 161.4]      \\
\addlinespace 
July 15, 2021   &  106.7 [60.4, 149.9]         & 99.8 [21.8, 181.1]        & 67.6 [-88.6, 206.4] &  4.1 [-39.5, 46.8]            \\
\addlinespace 
October 10, 2021 &    102.6 [92.1, 113.2]       & 81.5 [73.8, 89.5]        & 59.9 [50.1, 69.6] &  54.2 [36.7, 73.7]            \\
\addlinespace 
January 10, 2022 &   159.2 [150.8, 168.0]        &  160.5 [147.4, 174.1]       & 144.2 [119.7, 171.0] &    145.5 [130.0, 160.8]          \\
\addlinespace 
May 08, 2022 & 87.9 [76.7, 99.3]          &    59.8 [52.4, 67.5]     & 24.5 [17.0, 32.7] & 18.4 [-0.4, 37.9]             \\
\addlinespace 
May 24, 2022 &  -94.5 [-116.9, -71.3]         &  -80.8 [-101.0, -60.7]       & -67.6 [-86.6, -48.5] &     -49.8 [-95.6, 0.6]         \\
\addlinespace 
July 31, 2022 &   356.7 [326.9, 387.0]        &  364.1 [301.5, 427.5]       & 262.4 [188.0, 334.2] &  234.5 [166.6, 303.7]            \\
\addlinespace 
September 02, 2022 &  123.1 [115.2, 131.5]         &  109.8 [97.5, 121.8]       & 103.8 [85.7, 121.8]  &    77.5 [61.1, 93.6]          \\
\bottomrule
\end{tabular*}
}
\label{table:2}
\end{table}

\begin{table}[!ht]
\centering
\caption{Two sample KS test statistic values for the bootstrapped acceleration distribution for all the cadences of CME occurred on June 10, 2021.}
{\footnotesize
\begin{tabular*}{\textwidth}{
  @{\extracolsep{\fill}}
  l
  c
  c
  c
  c
  @{}
}
\toprule
  & \multicolumn{4}{c}{KS statistic}\\ \cline{2-5}\addlinespace
 \multicolumn{1}{c}{Cadence} & 30 s & 1 min & 2 min & 5 min \\
\cline{2-5}
   \multicolumn{1}{c}{30 s}       &  \multicolumn{1}{|c}{-}    &  0.68     &  0.85       &   0.88  \\
\multicolumn{1}{c}{1 min}          &\multicolumn{1}{|c}{0.68 }       &  - & 0.50          & 0.73  \\
\multicolumn{1}{c}{2 min}             & \multicolumn{1}{|c}{0.85}          & 0.50         &  -          &  0.44       \\
\multicolumn{1}{c}{5 min}   &  \multicolumn{1}{|c}{0.88 }         & 0.73         & 0.44  &  -            \\
\bottomrule
\end{tabular*}
}
\label{table:4}
\end{table}

\begin{figure}[ht]
\centering
\includegraphics[width =\textwidth]{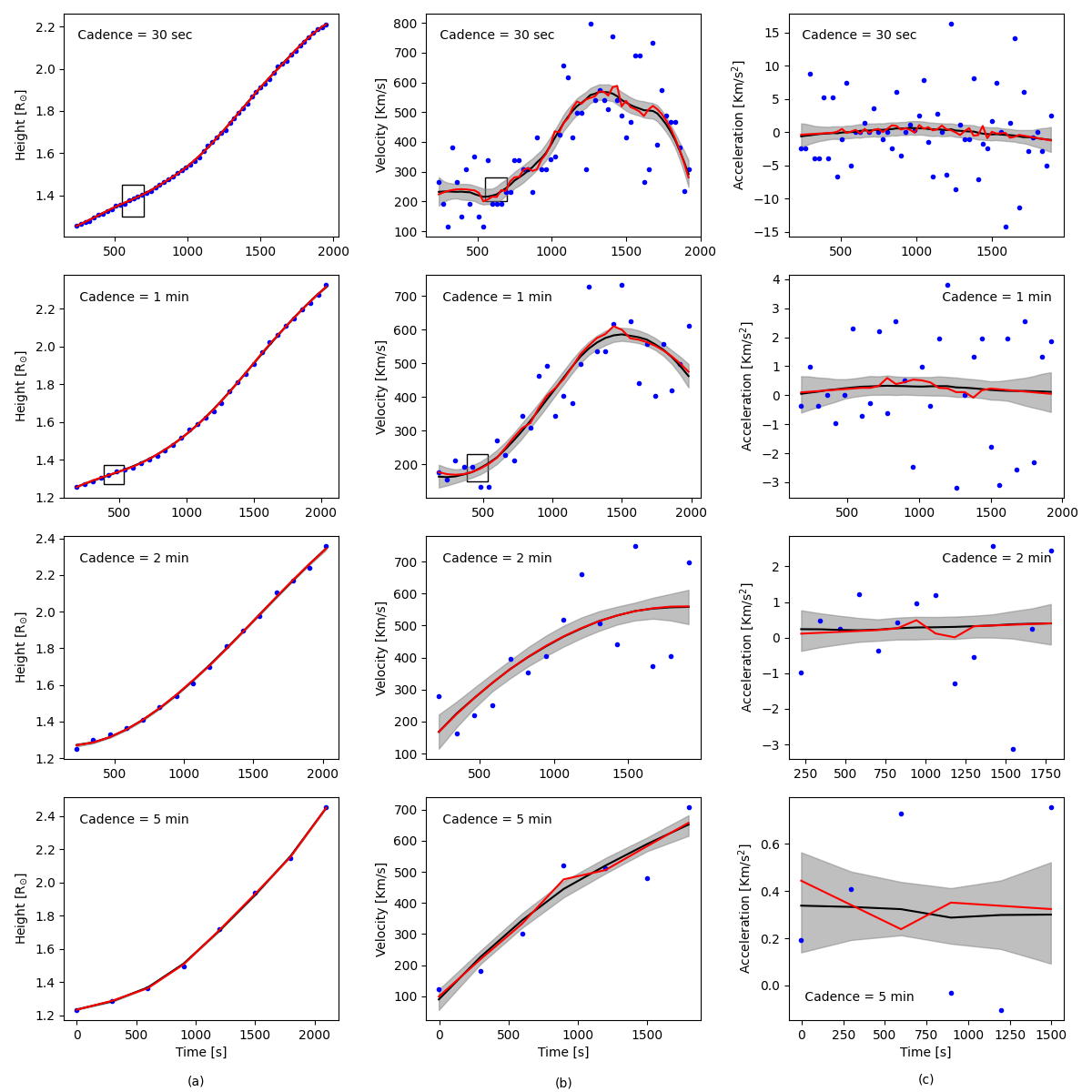}
\caption{A complete kinematic profile of the CME observed on June 10, 2021. (a) represents the bootstrapped Savitzky-Golay filter applied height-time plot for each cadence (from top to bottom). (b) and (c)  represents the corresponding velocity and acceleration plot, respectively. The blue dot represents (a) the height measurement taken using the K-Cor coronagraph and (b) and (c) corresponding numerically derivative velocity and acceleration. The red line represents the fitted curve. The black line represents the median value, and the grey-shaded region depicts the interquartile range. The rectangular box in the height and velocity time plot for 30 s and 1 min depicts the position of the knee in the kinematic profile.}
\label{fig:ht_vt_at}
\end{figure}
\begin{figure}[ht]
\includegraphics[width=\textwidth]{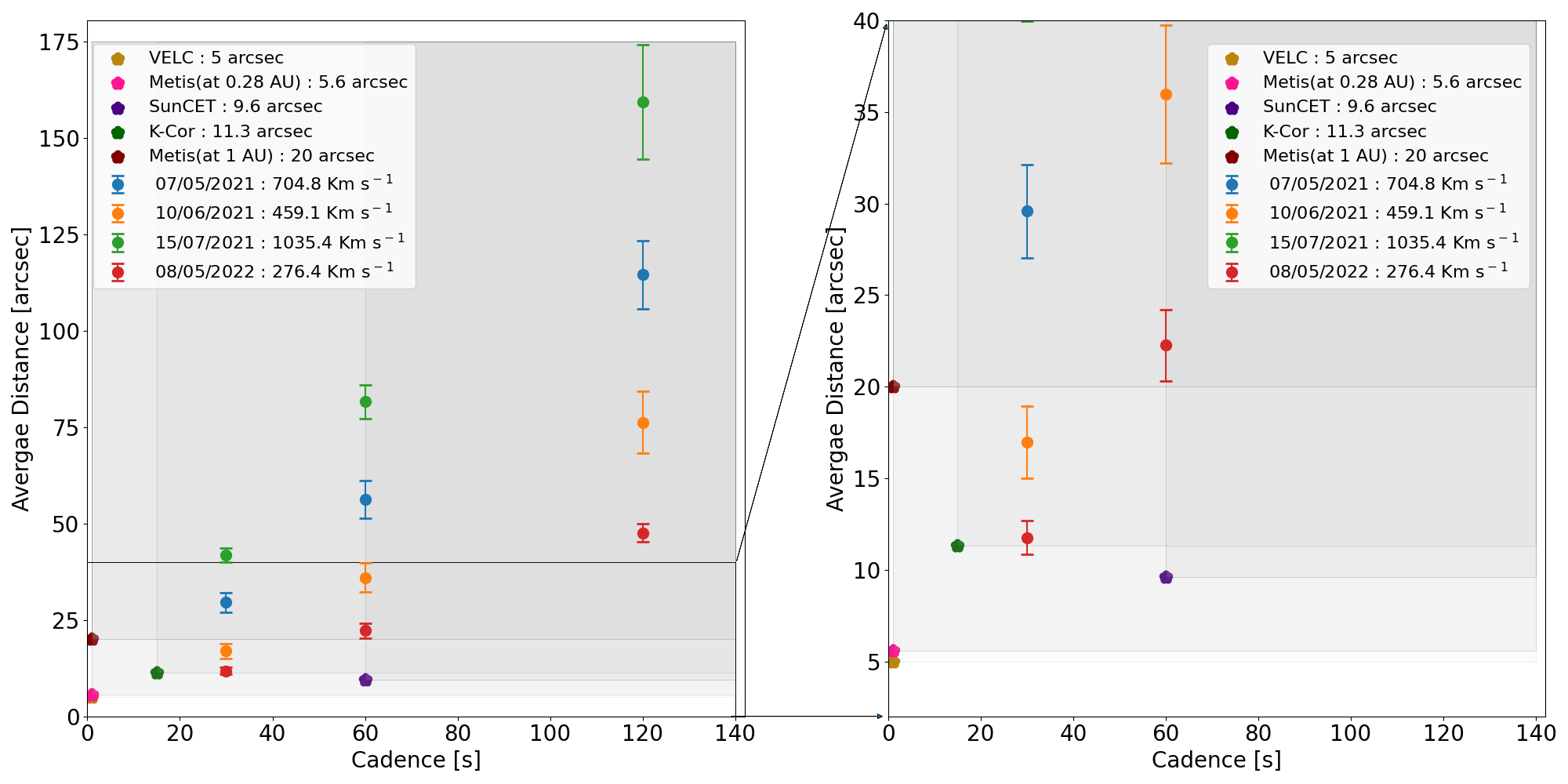}
\caption{ Average distance covered by the CME in arcseconds for different time intervals by different CMEs. The error bars represent the error in the mean of the distance covered by the CME. The position of horizontal lines represents the spatial resolution for different instruments, and the length of each line corresponds to the possible cadence of the instruments.  }
\label{fig:pixel_diff}
\end{figure}
The Quadratic fit assumes constant acceleration during the CMEs propagation. However, earlier studies have suggested that the acceleration does not remain constant, especially at lower heights \citep{Majumdar_2020}, and different initiation mechanisms show different kinematic profiles (see \cite{Mierla2013}). 
In Figure \ref{fig:ht_vt_at}, We plot the height-time data smoothed using the bootstrapped Savitzky-Golay filter in the first column for each cadence, represented by the red line. The second and third columns display the velocity and acceleration profiles, respectively. The red line represents the smoothed fit using the Savitzky-Golay filter for each cadence. With the help of bootstrapping technique, we calculated the median and interquartile range of the dataset. The grey shaded region represents the interquartile range of the data set. The black curve in each plot represents the median of the data set. We observe a change in the shape of the kinematic profile upon changing the cadence. Specifically, in the second column of Figure \ref{fig:ht_vt_at}, we noticed that the CME experiences a sudden increase in velocity (which marks the onset of the main acceleration phase) during propagation for 30 s and 1 min cadence. However, this feature is not visible in the degraded cadence data. To better understand this, we also mark the position of the knee (by a rectangular box) in the height-time profile, which indicates the onset of acceleration. We find that a cadence of 30 s and 1 min is able to capture this point of onset of acceleration (although these heights are slightly different for the two cadences) while degrading the cadence to 2 mins and 5 mins leads to washing out of this height. Hence, it becomes impossible to measure the height of the onset of acceleration. Certain CMEs exhibit similar behaviour, where we observe this impulsive acceleration phase for high cadence while this regime is not captured in low cadence observations (see supplementary material for more examples). On the other hand, for CMEs exhibiting a gradual increase in velocity over time, the change in cadence from 1 min to 2 min has minimal effect on their kinematics profile. This is expected because such gradual CMEs are prone to experiencing a small acceleration magnitude for a long duration, which makes them bereft of such knee in their height-time profile, thus reducing the importance of cadence in deciding the height of onset of acceleration. Another thing worth noting is that for most CMEs, regardless of whether they exhibit a gradual or a sudden increase in velocity, the velocity-time plot for the 30 s cadence data shows a significant amount of scatter. This scatter is likely a result of the difficulties and uncertainties involved in precisely tracking the leading edge of the CMEs within shorter time intervals. In order to resolve changes in the position of the CME's leading edge, a minimum height coverage of 2 pixels is required. This means that CME should move at least for 2 pixels to distinguish its position in subsequent images clearly. This distance corresponds to the spatial resolution of the instrument, and any movement below this threshold will not be discernible in the data. This height coverage corresponds to 11.3 arcseconds, considering the pixel scale of the K-Cor coronagraph.
In Figure \ref{fig:pixel_diff}, we plot the average height coverage in arcseconds for four different CMEs observed by the K-Cor coronagraph. We have carefully selected these CMEs based on their average velocities, aiming to cover a diverse range of CMEs with varying speeds. The chosen CMEs exhibit significant differences in their average velocities. Specifically, we have included 4 CMEs with speeds of 276.4, 459.1, 704.8, and 1035.4 Km/s to ensure a comprehensive representation of CMEs with different velocities. The average velocities of these CMEs were obtained through the linear fitting of the height-time plots (Table \ref{table:1}). We observe that CMEs with velocities below 500 km/s exhibited an average height coverage of less than 20 arcseconds in 30 s, which corresponds to 4 pixels based on the pixel scale of the K-Cor coronagraph. The slowest CME, with a velocity of 276 km/s, covered an average height of 11.4 arcseconds in 30 s, corresponding to approximately 2 pixels. It is noteworthy that even for 1 min cadence, the slowest CME only travelled an average distance of 4 pixels in 1 min. Thus, this clearly shows that tracking slow CMEs in high cadence might be challenging. 
However, the fastest CME, with a velocity of 1035.4 km/s, exhibited an average distance coverage of 7 pixels in 30 s, which makes it relatively easier to identify changes in the leading edge position in successive frames. However, manual tracking of such subtle changes often introduces uncertainties in the tracking. Further, it must also be noted that the spatial resolution of K-Cor at 30 s cadence poses challenges in accurately discerning changes in the CME's leading edge position, especially for slow CMEs. This issue remains relevant even when using 1 min cadence for slower CMEs. Therefore, arriving at a single optimum cadence for observing all CMEs might be tricky, as this work clearly shows that a successful tracking of a CME is largely dictated not only by the cadence of the instrument but also by the speed of the CME.\par
Existing and upcoming missions, such as Metis on board Solar Orbiter \citep[Metis; ][]{metis}, Visible Emission Line Coronagraph \citep[VELC; ][]{singh2011,Singh2013,raghavendra} on board Aditya-L1 \citep{seetha}, and The Sun Coronal Ejection Tracker \citep[SunCET;][]{Suncet}, are poised to improve our understanding of coronal mass ejections (CMEs) by capturing them with enhanced spatial and temporal resolution. Thus, making the best use of these resources requires effectively planning the observational campaigns with these missions, which is demonstrated in Figure \ref{fig:pixel_diff}. The differently shaded regions in Figure \ref{fig:pixel_diff} represent the spatial and temporal coverage of each instrument discussed above. Each shaded region is marked with an asterisk in a different colour, indicating the highest achievable spatial and temporal resolution at the bottom left vertex of the respective region. There is a clustering of events and close overlap of different mission regimes from 0 to 20 arcsec in the left plot of Figure \ref{fig:pixel_diff}. Thus, In order to provide a more clear picture of different regimes of different missions, we plot a zoomed-in version of the left plot of Figure \ref{fig:pixel_diff} in the right plot. The positions of the data points in this plot also provide insights into the optimal cadence for tracking different CMEs using different instruments. The Metis and VELC coronagraphs offer impressive capabilities in terms of both spatial and temporal resolution. Metis is currently en route to the Sun and will approach as close as 0.28 AU from it. As the distance between Metis and the Sun varies during its journey, its spatial resolution will also change, peaking at 5.6 arcseconds when it reaches the closest perihelion of 0.28 AU. On the other hand, VELC offers a spatial resolution of 5 arcseconds. Both instruments can achieve a temporal resolution of 1 second. Taking into account a CME observed on June 10, 2021, with an average speed of 459 Km s$^{-1}$ (taken from Table \ref{table:1}), we find that for a cadence of 30 s, the spatial resolution needs to be much better than 15 arcseconds, which is possible for Metis and VELC, but not for SunCET (or K-Cor, as K-Cor just fits in with a resolution of 11.3 arcseconds), thus using a cadence of 1 min can be more effective for such CMEs. On the other hand, for a slow CME, the situation is more tricky, as the spatial resolution needs to be better than 11 arcseconds for a cadence of 30 seconds. This is again possible with VELC and Metis but not with K-Cor and SunCET. However, with a cadence of 1 min, it is possible to track the CME using SunCET, VELC, and K-Cor, but it might not be possible with Metis (when it is at 1 AU). However, for very fast CMEs, with speeds higher than 700 Km/s, 30 s cadence could be useful in capturing their main acceleration phase, provided the CME is bright enough to get tracked in successive frames, while the other cases, except for very slow CMEs with speeds lesser than 300 Km/s, a cadence of 1 min seems to be at the optimum level. But, it is essential to note that manual tracking of CMEs done at such a high cadence where you resolve the CME in the subsequent frames only through a couple of pixels can introduce errors, and it requires caution when interpreting the results obtained from such observations.
\section{Summary and Conclusion}
\label{sec-summary}
We conducted a study to investigate the impact of imaging instrument cadence on the kinematic profile of a CME. As previously mentioned, understanding the kinematic profile can provide insights into the underlying initiation mechanism of CMEs and provide crucial insights on their evolution, and the construction of this kinematic profile is largely limited by the cadence of the instrument. Thus, we performed a study to understand the impact of image cadence on CME kinematics. \\ 
We examined 10 CMEs and calculated their average velocity and acceleration. The leading edge, being a diffuse structure, can introduce additional uncertainty in height measurements. To minimize this uncertainty, we repeated the process five times and reported the mean and standard deviation as the height measurement of the leading edge, along with the measurement error (Figure \ref{fig:fiting}). To obtain the average velocity and acceleration, we used the bootstrapping technique while fitting the linear and quadratic profiles to the height-time plots of these CMEs. This technique allowed us to obtain different confidence intervals for the average speeds and accelerations corresponding to different data cadences, thus showing the dependence of the former on the latter (Figure \ref{fig:boot_vel_acc}). It seems with the degradation of image cadence, the confidence interval goes through notable change for the case of average acceleration, while the change in the case of average velocities is not that pronounced. To understand the impact of change in cadence on the velocities and accelerations, we observed that the average velocity values obtained from linear fitting did not exhibit significant changes with variation in the cadence of the observations (Table \ref{table:1}). We noticed the change in the average velocity is within the 12\% of its value for 30 s cadence (Fig \ref{fig:abs_rel_vel}(a)). However, we observed a significant dependency of the average acceleration on the cadence of observations when we used the quadratic fit for the height-time plot (Table \ref{table:2}). The average acceleration exhibited a significant change of more than 70\% compared to its value for 30 s cadence for most of the CMEs (Figure \ref{fig:abs_rel_acc}(b)). \\
Keeping in mind that CMEs do not strictly propagate with constant acceleration, it has been pointed out by many authors \citep{james_chen, Zhang_2001, Zhang_2004, Zhang_2006, Bein_2011, Majumdar_2020}that the use of a constant acceleration model may not accurately capture the true nature of CME kinematics. To address this issue, we used the Savitzky-Golay filter, which provides a smoothed kinematic profile while preserving the underlying characteristics of the CMEs. We observed a notable variation in the kinematic profile of CMEs when altering the observation cadence. CMEs exhibiting an impulsive increase in velocity were particularly affected by changes in cadence, as with the degradation of the cadence, the time (and/or height) of onset of acceleration could not be traced, and the acceleration phase gets diluted (Figure \ref{fig:ht_vt_at}). On the other hand, the kinematic profile of gradual CMEs remains less affected by changes in cadence, as they do not experience any such impulsiveness in their acceleration.\\ 
Our investigation revealed that not only does the temporal resolution play a role in the tracking process, but the spatial resolution of the instrument also has a significant impact, especially for very high cadence data. The spatial resolution directly influences the detectability of position changes in the targeted CME feature across successive images. When the spatial resolution is low, these position changes become indistinguishable, while with high cadence and manual tracking, such small changes introduce uncertainties leading to a significant scatter in the derived velocity-time plot (Figure \ref{fig:ht_vt_at}). We also find that in addition to considering the temporal and spatial resolution of the instrument, the velocity of the CME also plays a significant role in the successful tracking of them. Hence, determining a single optimal cadence to accurately observe and track all CMEs can be challenging. \\ 
Thus, in a nutshell, it seems, to track CMEs ranging from very slow (speeds less than 300 Km/s) to very fast (speeds greater than 700 Km/s), a cadence of 1 min with a pixel resolution $\sim$5 arcseconds can be good enough for confident tracking and successful capturing of their evolutionary phases, which would be possible with VELC and Metis. In this regard, it should also be noted that the overlapping of the differently shaded regions indicates the possibility of having combined and complementary observations in future to study the kinematics of CMEs travelling at different speeds. In this regard, it is also worth noting that knowing a priori if a CME is going to be fast or slow is not straightforward, but given the understanding that CMEs coming from the quiet Sun regions are more prone to be gradual CMEs, while the ones coming from energetic active regions can be impulsive and fast CMEs, the disk observations (for example from SUIT on board Aditya-L1  \citep[SUIT; ][]{Tripathi_2017} can be used to identify the prevalence of potential pre-eruptive features and the cadence of the observational plans could be decided accordingly. These results should also be taken into consideration when working with automated CME detection algorithm \citep{Patel2018}. We believe this study will aid in the planning of observational campaigns with existing and upcoming coronagraphs, and help improve our current understanding of CME evolution in the inner corona. \par
\section*{Conflict of Interest Statement}
The authors declare that the research was conducted in the absence of any commercial or financial relationships that could be construed as a potential conflict of interest.

\section*{Author Contributions}

NV was responsible for leading the analysis, including image processing and all the presented analyses. The events for the analysis were identified by NV, SM, and RP. SM, RP, VP, and DB assisted in interpreting the results. SM, RP, VP, and DB also aided in structuring the draft and in language correction. NV prepared the manuscript, and all authors contributed to the discussions.

\section*{Funding}
NV is supported by funds of the Council of Scientific \& Industrial Research (CSIR), India, under file no.
09/0948(0007)/2020-EMR-I

\section*{Acknowledgments}
 We thank the anonymous referees for their valuable comments which have improved the manuscript. The authors express their gratitude to ARIES for providing us with the necessary computational resources. We also acknowledge the Mauna Loa Solar Observatory, operated by the High Altitude Observatory, as part of the National Center for Atmospheric Research (NCAR) for providing us with the required data. The authors would like to acknowledge the support provided by the National Science Foundation to NCAR. The SECCHI data used here were produced by an international consortium of the Naval Research Laboratory (USA), Lockheed Martin Solar and Astrophysics Lab (USA), NASA Goddard Space Flight Center (USA), Rutherford Appleton Laboratory (UK), University of Birmingham (UK), Max-Planck-Institut for Solar System Research (Germany), Centre Spatiale de Liège (Belgium), Institut d'Optique Théorique et Appliquée (France), Institut d'Astrophysique Spatiale (France). We also acknowledge SDO team for making AIA data available and SOHO team for LASCO data. The authors would also like thank Prof. Prasad Subramanian for the initial discussions with him.

\section*{Supplemental Data}
 File attached.

\section*{Data Availability Statement}
The data utilized in this study can be accessed from the following data archives:
 \href{https://mlso.hao.ucar.edu/mlso_data_calendar.php?calinst=kcor}{ MLSO K-COR DATA SUMMARY.}
\bibliographystyle{Frontiers-Harvard}
\bibliography{bibliography}
\end{document}